\documentclass[a4paper]{article}
\usepackage{ifxetex}
\ifxetex
    \usepackage{gc2019eng}
\else
    \usepackage{gc2019eng_latex}
\fi
\usepackage{hyperref}
\usepackage{subcaption}
\usepackage{graphicx} 

\title{Hacking VMAF with Video Color and Contrast Distortion}

\author{A.~Zvezdakova$^1$, S.~Zvezdakov$^2$, D.~Kulikov$^3$, D.~Vatolin$^4$}

\email{\{azvezdakova|szvezdakov|dkulikov|dmitriy\}@graphics.cs.msu.ru}

\organization{$^{1,2,4}$Lomonosov Moscow State University, Moscow, Russia;\par $^3$Lomonosov Moscow State University, Moscow, Russia; Dubna State University, Dubna, Russia}

\abstract{Video quality measurement takes an important role in many applications. Full-reference quality metrics which are usually used in video codecs comparisons are expected to reflect any changes in videos. In this article, we consider different color corrections of compressed videos which increase the values of full-reference metric VMAF and almost don't decrease other widely-used metric SSIM. The proposed video contrast enhancement approach shows the metric inapplicability in some cases for video codecs comparisons, as it may be used for cheating in the comparisons via tuning to improve this metric values.}

\keywords{video quality, quality measuring, video-codec comparison, quality tuning, reference metrics, color correction.}

\authorsInfo
{
    Anastasia Zvezdakova, PhD student. Received her M.S. degree in computer science from Moscow State University in 2018. Currently, she is a PhD student at Moscow State University and a member of Video Group in MSU Graphics\&Media Lab. Her research interests involve video codecs analysis and optimization, stereoscopic video subjective quality assessment. Anastasia is one of the contributors to MSU Video Codec Comparison Project \url{http://www.compression.ru/video/codec_comparison/index_en.html} and to the 3D video quality measurement project VQMT3D \url{http://compression.ru/video/vqmt3d/}. E-mail aantsiferova@graphics.cs.msu.ru

    Sergey Zvezdakov, PhD student. Received his M.S. degree in computer science from Moscow State University in 2018. Currently, he is a PhD student at Moscow State University and a member of Video Group in MSU Graphics\&Media Lab. His research interests involve video codecs analysis and optimization. Sergey is one of the contributors to MSU Video Codec Comparison Project \url{http://www.compression.ru/video/codec_comparison/index_en.html. E-mail szvezdakov@graphics.cs.msu.ru}

    Dmitriy Kulikov, PhD in computer science. Received his M.S. degree in 2005 and his PhD in 2009, both from Moscow State University. Currently, he is head of the MSU Video Codec Comparison Project in Video Group at the CS MSU Graphics\&Media Lab. Also, he is an associate professor in Dubna State University. His research interests include compression methods, video processing algorithms, video quality metrics, machine learning. Dmitriy is one of the main contributors to MSU Video Codec Comparison Project \url{http://www.compression.ru/video/codec_comparison/index_en.html}.  E-mail dkulikov@graphics.cs.msu.ru

    Dmitriy Vatolin, PhD in computer science, head of Video Group (CS MSU Graphics\&Media Lab). Graduated from the Computer Science department of Lomonosov Moscow State University in 1996 and got his PhD in 2000. The theme of his PhD thesis was ``Optimization methods of fractal image compression''. Dmitriy has been teaching a computer graphics course at MSU since 1997. He wrote the book Algorithms of Image Compression in 1999 and coauthored Methods of Data Compression in 2003. Dr Vatolin supervised collaborative research projects with Intel, Real Networks and Samsung. He created one of the largest websites devoted to data compression (http://compression.ru). He teaches courses on methods of 3D and 2D video and image processing and compression. E-mail dmitriy@graphics.cs.msu.ru
}

\begin{document}

\maketitle

\begin{multicols*}{2}

\section{Introduction}

At the moment, video content takes a significant part of worldwide network traffic and its share is expected to grow up to 71\% by 2021 \cite{cisco_2016}. Therefore, the quality of encoded videos is becoming increasingly important, which leads to growing of an interest in the area of new video quality assessment methods development. As new video codec standards appear, the existing standards are being improved. In order to choose one or another video encoding solution, it is necessary to have appropriate tools for video quality assessment. Since the best method of video quality assessment is a subjective evaluation, which is quite expensive in terms of time and cost of its implementation, all other objective methods are improving in an attempt to approach the ground truth-solution (subjective evaluation).

Methods for evaluating encoded videos quality can be divided into 3 categories \cite{Chikkerur}: full-reference, reduced-reference and no-reference. Full-reference metrics are the most common, as their results are easily interpreted --- usually as an assessment of the degree of distortions in the video and their visibility to the observer. The only drawback of this approach compared to the others is the need to have the original video for comparison with the encoded, which is often not available.

\begin{figure}[H]
    \centering{\includegraphics[width=\linewidth]{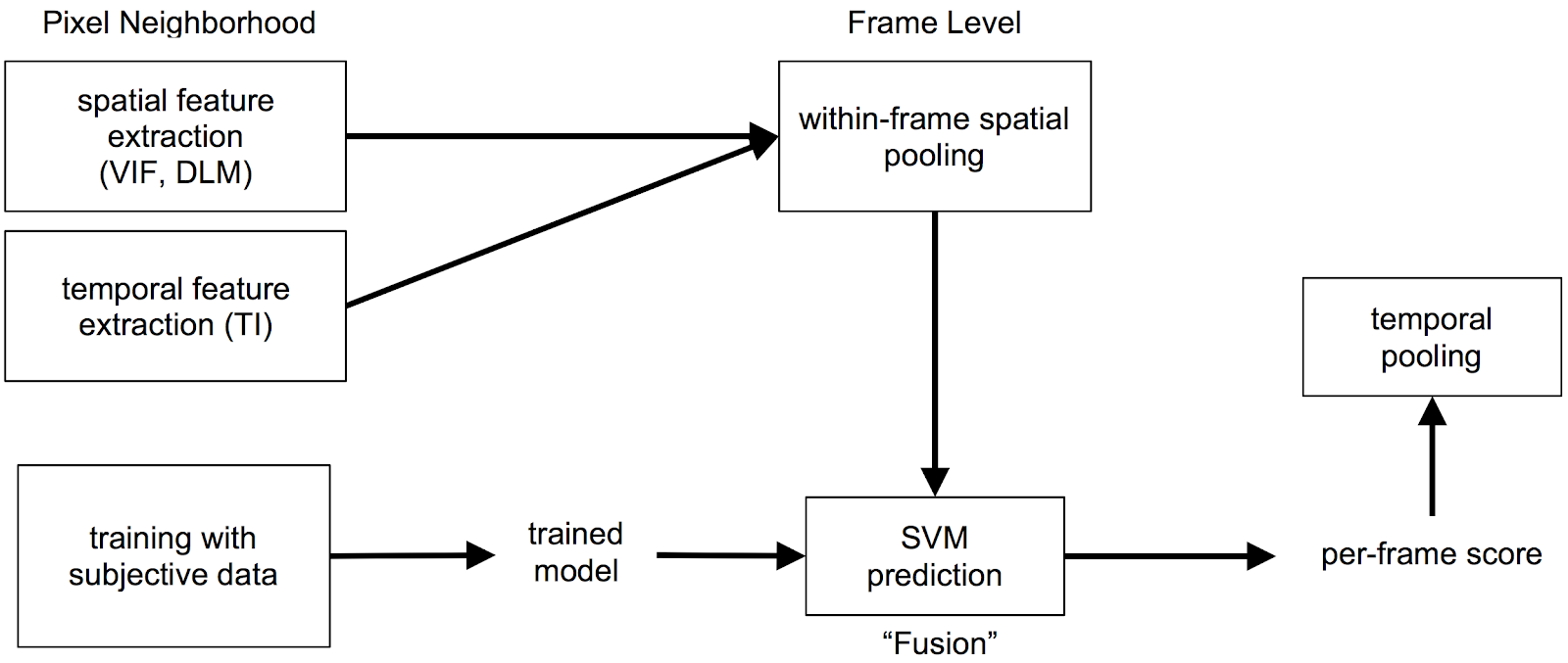}}
    \caption{The scheme of VMAF algorithm.}
    \label{fig:scheme}
\end{figure}

One of the widely-used full-reference metrics which is gaining popularity in the area of video quality assessment is Video Multimethod Fusion Approach (VMAF)\cite{vmaf_article}, announced by Netflix. It is an open-source learning-based solution. Its main idea is to combine multiple elementary video quality features, such as Visual Information Fidelity (VIF)\cite{vif}, Detail Loss Metric (DLM)\cite{dlm} and temporal information (TI) -- the difference between two neighboring frames, and then to train support vector machine (SVM) regression on subjective data. The resulting regressor is used for estimating per-frame quality scores on new videos. The scheme \cite{Bampis} of this metric is shown in Fig.~\ref{fig:scheme}.

Despite increasing attention to this metric, many video quality analysis projects, such as Moscow State University's (MSU) Annual Video Codec Comparison \cite{msu_2018}, still use other common metrics developed many years ago, such as structural similarity (SSIM) and even peak signal-to-noise ratio (PSNR), which are based only on difference characteristics of two images. At the same time, many readers of the reports of these comparisons send requests to use new metrics of VMAF type. The main obstacle for the full transition to the use of VMAF metrics is non-versatility of this metric and not fully adequate results on some types of videos \cite{ittiam_critic}. 

The main goal of our investigation was to prove the no-universality of the current version of VMAF algorithm. In this paper, we describe video color and contrast transformations which increase VMAF-score with keeping SSIM score the same or better. The possibility to improve full-reference metric score after adding any transformations to distorted image means that the metric can be cheated in some cases. Such transformations may allow competitors, for example, to cheat in video codecs comparisons, if they ``tune'' their codecs for increasing VMAF quality scores. Types of video distortions that we were looking for change the visual quality of the video, which should lead to a decrease in the value of any full-reference metric. The fact that they lead to an increase in the value of VMAF, is a significant obstacle to using VMAF for all types of videos as the main quality indicator and proves the need of modification of the original VMAF algorithm.

\section{Study Method}
During testing of VMAF algorithm for video codecs comparisons, we noticed that it reacts on contrast distortions, so we chose color and contrast adjustments as basic types of the searched video transformations. Two famous and common approaches for color adjustments were tested to find the best strategy for VMAF scores increasing. Two cases of transformations application to the video were tested: applying transformation before and after video encoding. In general, there was no significant difference between these options, because the compression step can be omitted for increasing VMAF with color enhancement. Therefore, further we will describe only the first case with adjustment before compression, and we leave the compression step because in our work VMAF tuning is considered in case of video-codec comparisons.

We chose 4 videos which represent different spatial and temporal complexity \cite{google_si_ti}, content and contrast to test transformations which may influence VMAF scores. All videos have FullHD resolution and high bit rate. \textit{Bay time-lapse} and \textit{} were filmed in flat colors, which usually require color post-processing. Three of the videos (\textit{Crowd run}, \textit{Red kayak} and \textit{Speed bag}) were taken from open video collection on media.xiph.org and one was taken from MSU video collection used for selecting testing video sets for annual video codecs comparison \cite{msu_2018}. The description (and sources) of the first three videos can be found on site \cite{xiph_media}, and the rest \textit{Bay time-lapse} video sequence contained a scene with water and grass and the grass and waves on the water.

Three versions of VMAF were tested: 0.6.1, 0.6.2, 0.6.3. The implementations of all three metric versions from MSU Video Quality Measurement Tool \cite{vqmt} were used. The results did not differ much, so the following plots are presented for the latest (0.6.3) version of VMAF.

\section{Proposed Tuning Algorithm}

For color and brightness adjustment, two known and widespread image processing algorithms were chosen: unsharp mask and histogram equalization. We used the implementations of these algorithms which are available in open-source scikit-image \cite{skimage} library. In this library, unsharp mask has two parameters which influence image levels: radius (the radius of Gaussian blur) and amount (how much contrast was added at the edges). For histogram equalization, a parameter of clipping limit was analyzed.
In order to find optimal configurations of equalization parameters, a multi-objective optimization algorithm NSGA-II \cite{nsga_2} was used. Only the limits for the parameters were set to the genetic algorithm, and it was applied to find the best parameters for each testing video.

\begin{figure}[H]
    \begin{minipage}[b]{.48\linewidth}
    \centering{\includegraphics[width=\linewidth]{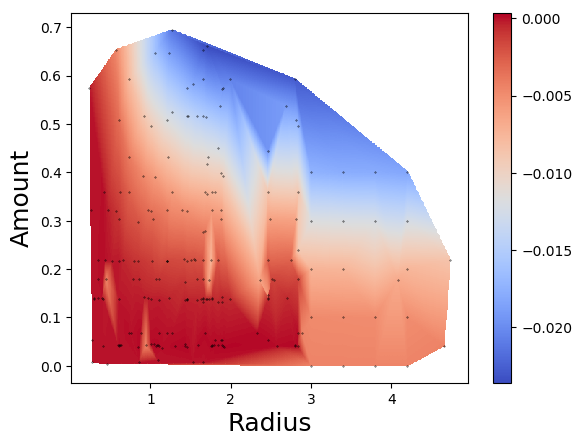}}
    \caption{SSIM scores for different parameters of unsharp mask on \textit{Bay time-lapse} video sequence.}
    \label{fig:ssim_bay}
    \end{minipage}
    \begin{minipage}[b]{.48\linewidth}
    \centering{\includegraphics[width=\linewidth]{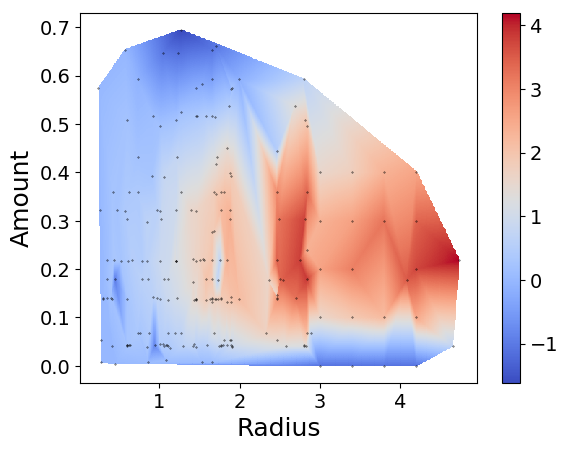}}
    \caption{VMAF scores for different parameters of unsharp mask on \textit{Bay time-lapse} video sequence.}
    \label{fig:vmaf_bay}
    \end{minipage}
\end{figure}

SSIM and VMAF scores were calculated for each video processed with the considered color enhancement algorithms with different parameters. As it was mentioned before, after color correction the videos were compressed with medium preset of x264 encoder on 3 Mbps. Then, the difference between metric scores of processed videos and original video were calculated to compare, how color corrections influenced quality scores. Fig.~\ref{fig:ssim_bay} shows this difference for SSIM metric of \textit{Bay time-lapse} video sequence for different parameter values of unsharp mask algorithm. The similarity scores for VMAF quality metric are presented in Fig.~\ref{fig:vmaf_bay}.

\begin{figure}[H]
    \centering{\includegraphics[width=.8\linewidth]{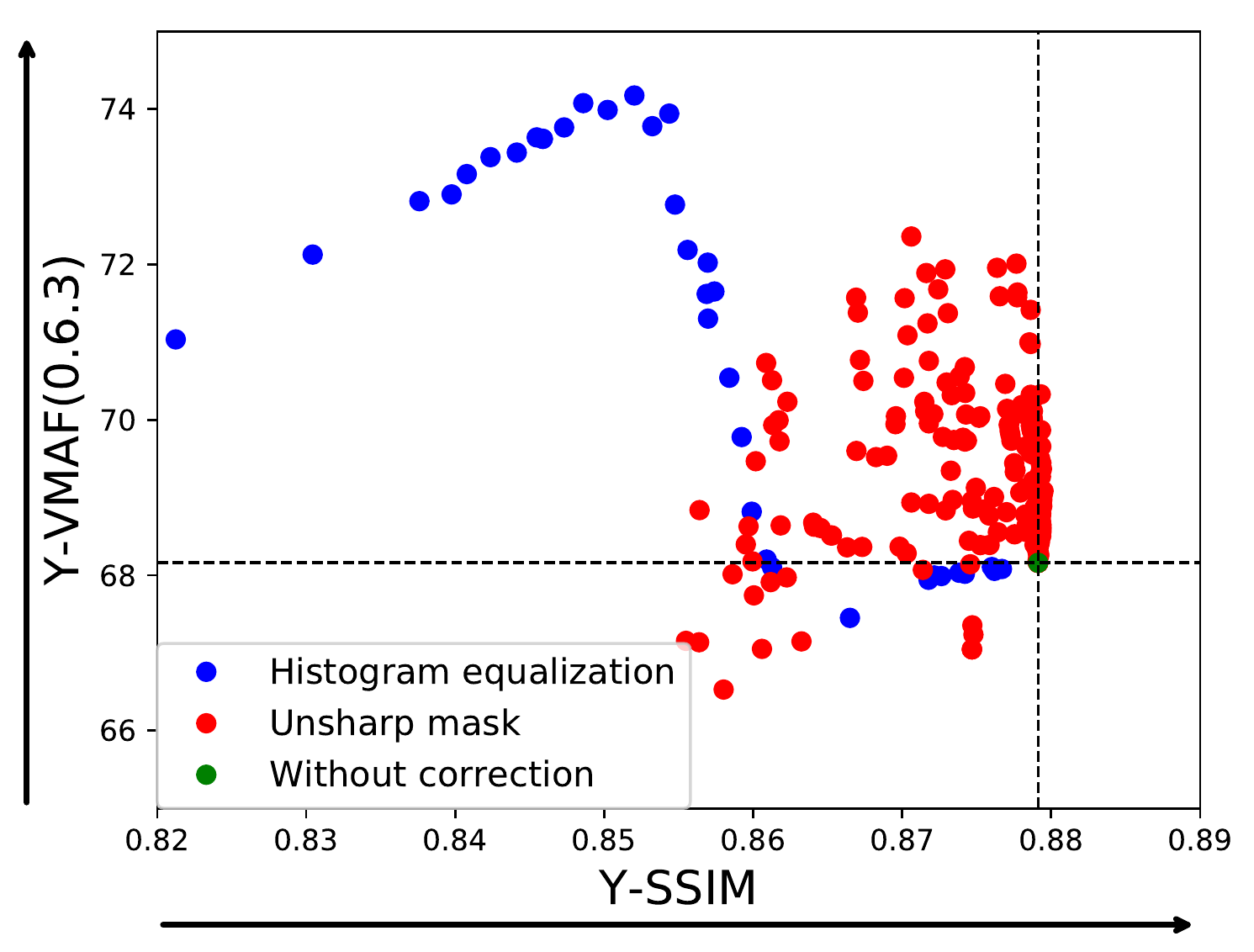}}
    \caption{Comparison of VMAF and SSIM scores for different configurations of unsharp mask and histogram equalization on \textit{Bay time-lapse} video sequence. The results in the second quadrant, where SSIM values weren't changed and VMAF values increased, are interesting for us.}
    \label{fig:vmaf_ssim_bay}
\end{figure}

On these plots, higher values mean that the objective quality of the color-adjusted video was better according to the metric. VMAF shows better scores for high radius and a medium amount of unsharp mask, and SSIM becomes worse for high radius and high amount. The optimal values of the algorithm parameters can be estimated on the difference in these plots. For another color adjustment algorithm (histogram equalization), one parameter was optimized and the results are presented on Fig.~\ref{fig:vmaf_ssim_bay} together with the results of unsharp mask. 

\begin{figure}[H]
    \begin{subfigure}{.48\textwidth}
    \centering{\includegraphics[width=.8\linewidth]{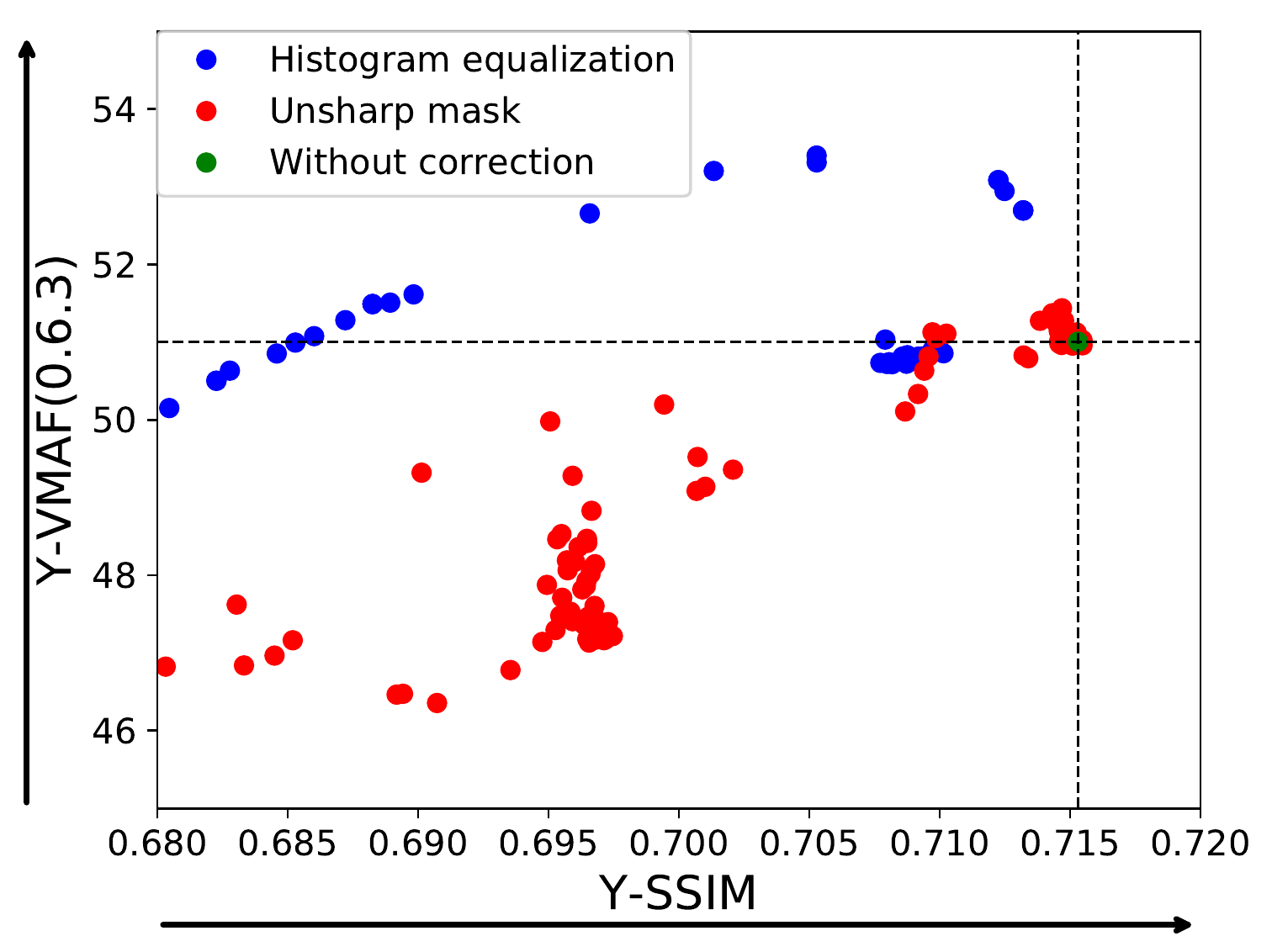}}
    \caption{Color tuning results for \textit{Crowd run} video sequence.}
    \label{fig:vmaf_ssim_crowd}
    \end{subfigure}

    \begin{subfigure}{.48\textwidth}
    \centering{\includegraphics[width=.8\linewidth]{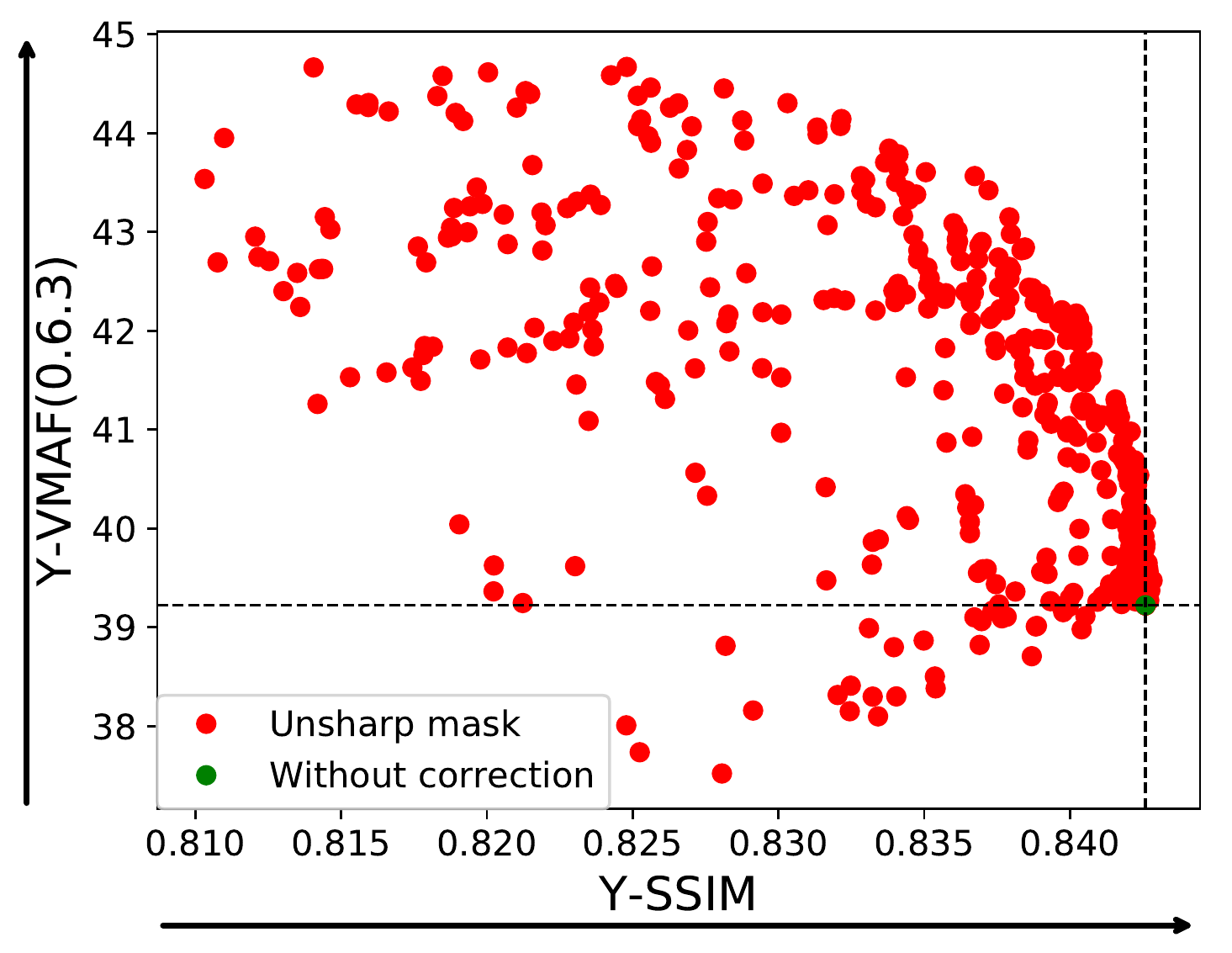}}
    \caption{Color tuning results for \textit{Red kayak} video sequence.}
    \label{fig:vmaf_ssim_red}
    \end{subfigure}

    \begin{subfigure}{.48\textwidth}
    \centering{\includegraphics[width=.8\linewidth]{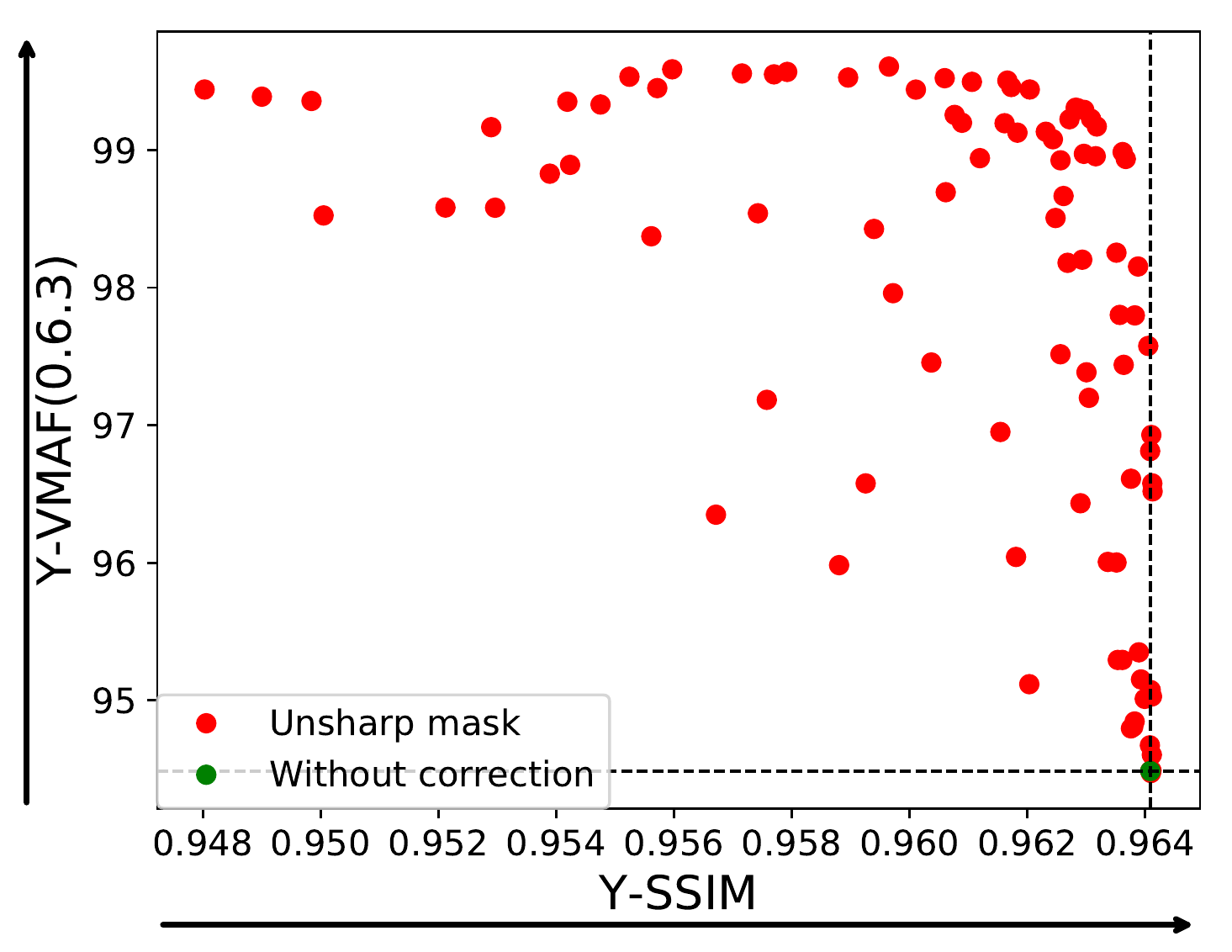}}
    \caption{Color tuning results for \textit{Speed bag} video sequence.}
    \label{fig:vmaf_ssim_speed}
    \end{subfigure}
    \caption{Comparison of VMAF and SSIM scores for different configurations of unsharp mask and histogram equalization on tested video sequences. The results in the second quadrant, where SSIM values weren't changed and VMAF values increased, are interesting for us.}
\end{figure}

According to these results, for some configurations of histogram equalization VMAF become significantly better (from 68 to 74) and SSIM doesn't change a lot (decrease from 0.88 to 0.86). The results slightly differ for other videos. On \textit{Crowd run} video sequence, VMAF was not increased by unsharp mask (Fig.~\ref{fig:vmaf_ssim_crowd}) and was increased a little by histogram equalization. For \textit{Red kayak} and \textit{Speed bag} videos, unsharp mask could significantly increase VMAF and just slightly decrease SSIM (Fig.~\ref{fig:vmaf_ssim_red} and Fig.~\ref{fig:vmaf_ssim_speed})

\section{Results}

\begin{figure}[H]
    \begin{subfigure}{.23\textwidth}
    \begin{center}
    \includegraphics[width=\linewidth]{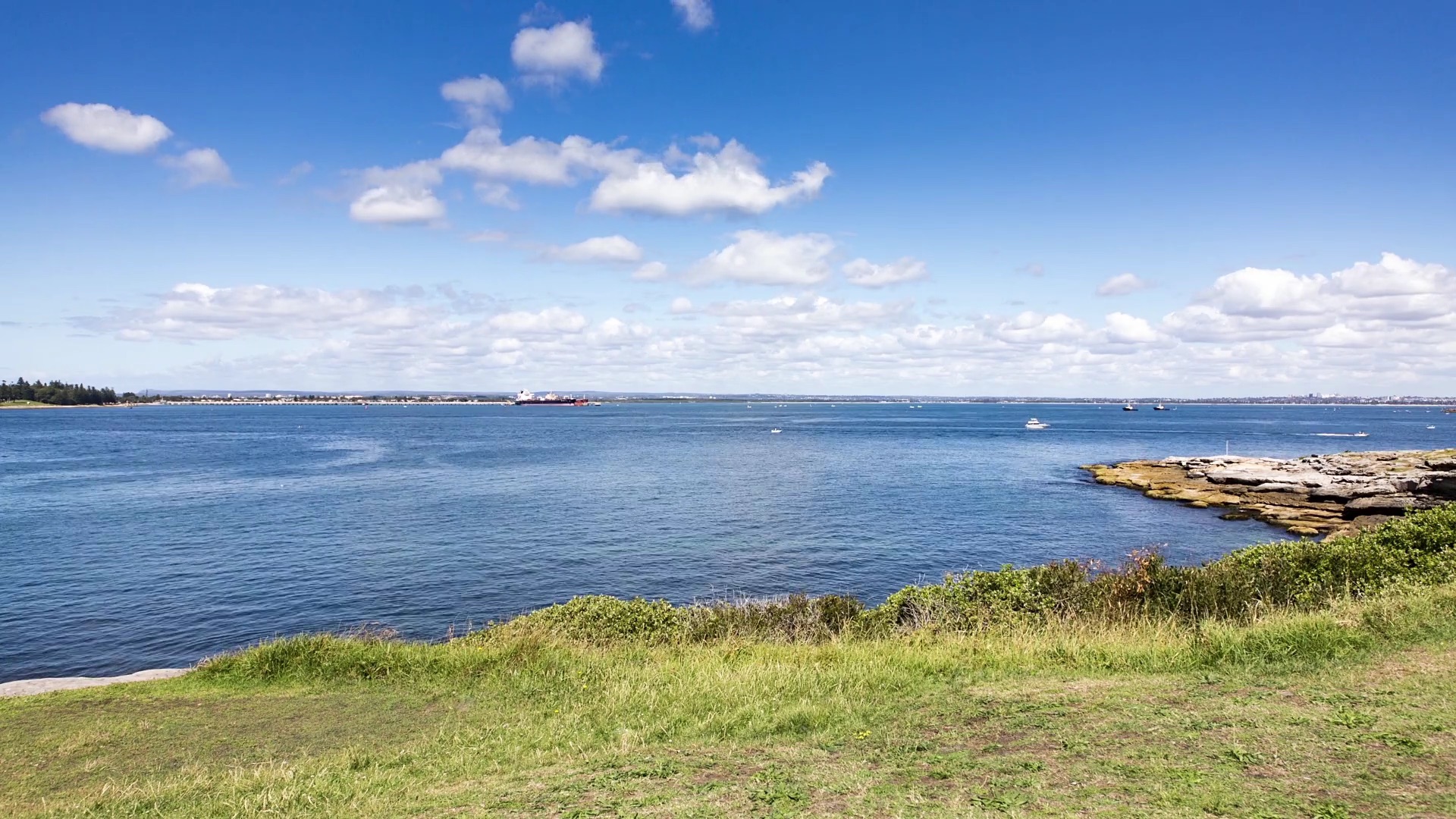}
    \\
    \includegraphics[width=\linewidth]{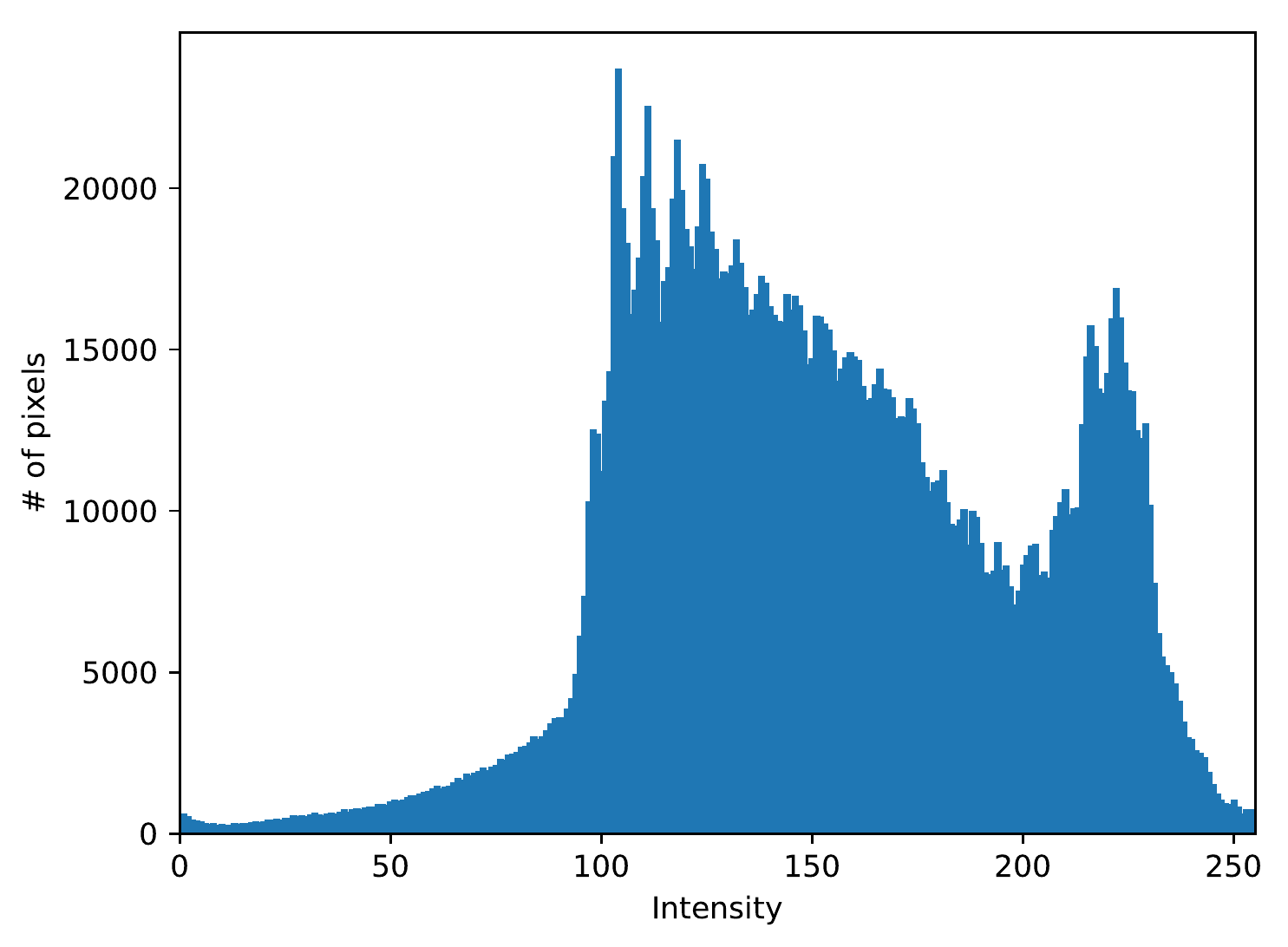}
    \end{center}
    \caption{Before unsharp mask\\ $VMAF = 68.160$, \\$SSIM = 0.879$}
    \label{fig:bay_time_lapse}
    \end{subfigure}
    \hspace*{\fill} 
    \begin{subfigure}{0.23\textwidth}
    \begin{center}
    \includegraphics[width=\linewidth]{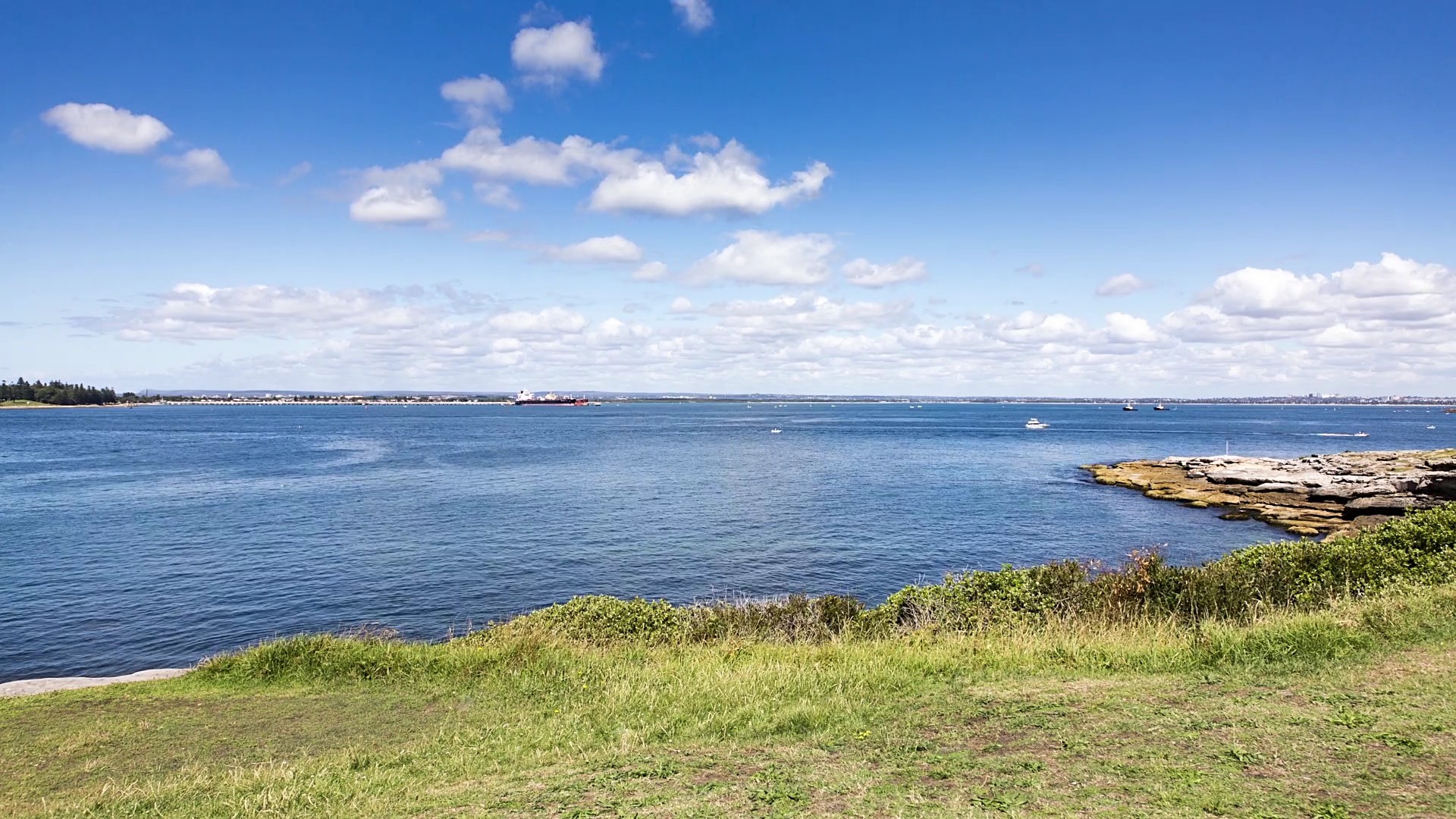}
    \\
    \includegraphics[width=\linewidth]{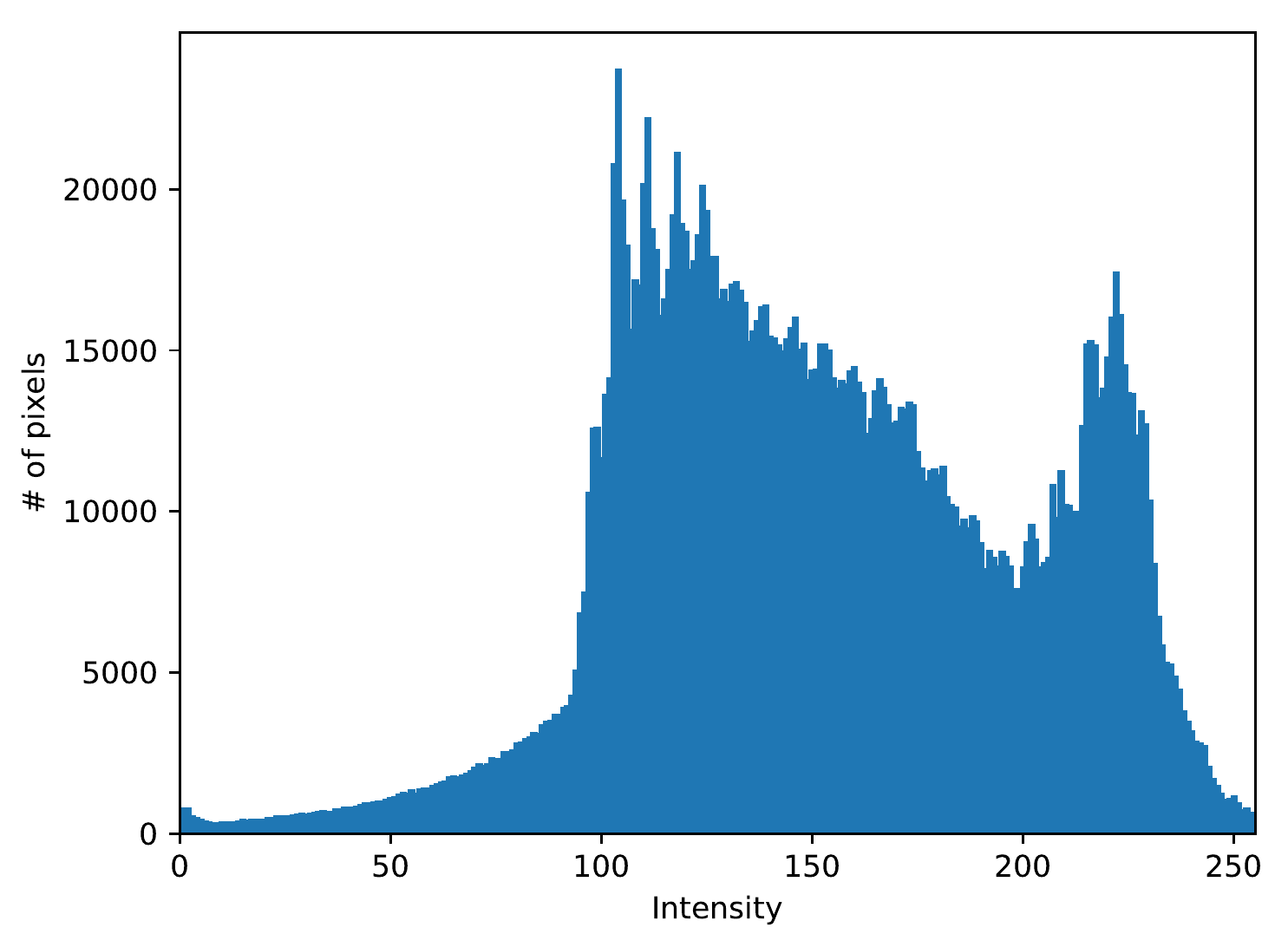}
    \end{center}
    \caption{After unsharp mask\\ $VMAF = 72.009$, \\$SSIM = 0.878$}
    \label{fig:bay_time_lapse_adj}
    \end{subfigure}
    \caption{Frame 5 from \textit{Bay time-lapse} video sequence and its histogram with and without contrast correction. Two images and their histograms look equivalent.}
\end{figure}

\begin{figure}[H]
    \includegraphics[width=\linewidth]{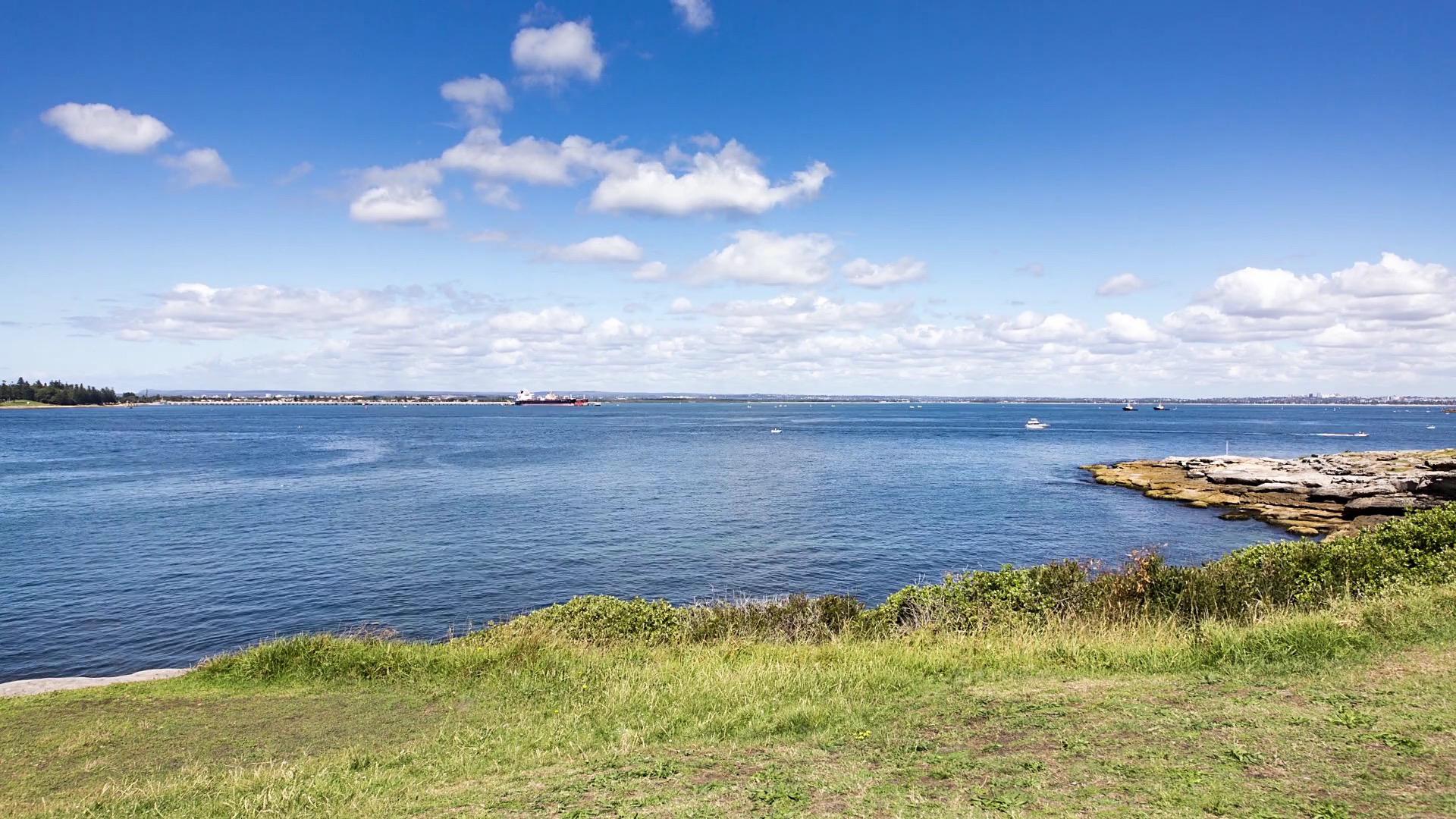}
    \caption{Checkerboard comparison of frame 5 from \textit{Bay time-lapse} video sequence before and after distortions. Two images look almost equivalent.}
    \label{fig:bay_timelapse_comparison}
\end{figure}

The following examples of frames from the testing videos demonstrate color corrections which increased VMAF and almost did not influence the values of SSIM. Unsharp mask with $radius = 2.843$ and $amount = 0.179$ increased VMAF without significant decrease of SSIM for \textit{Bay time-lapse} (Fig.~\ref{fig:bay_time_lapse} and Fig.~\ref{fig:bay_time_lapse_adj}). The images before and after masking look equivalent (a comparison in a checkerboard view is in Fig.~\ref{fig:bay_timelapse_comparison}) and have similar SSIM score, while VMAF score is better after the transforamion.

\begin{figure}[H]
    \begin{subfigure}{.23\textwidth}
    \begin{center}
    \includegraphics[width=\linewidth]{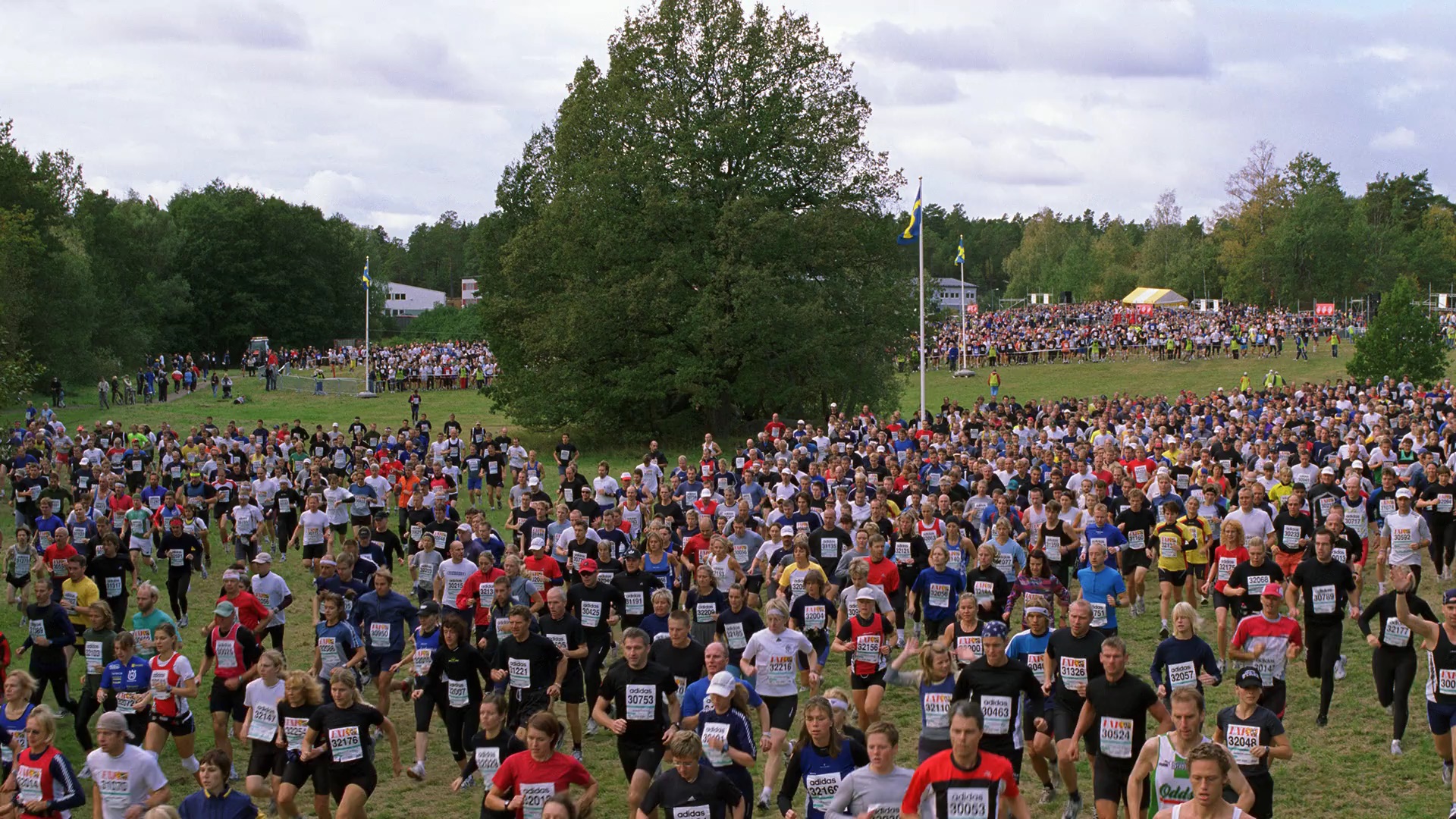}
    \\
    \includegraphics[width=\linewidth]{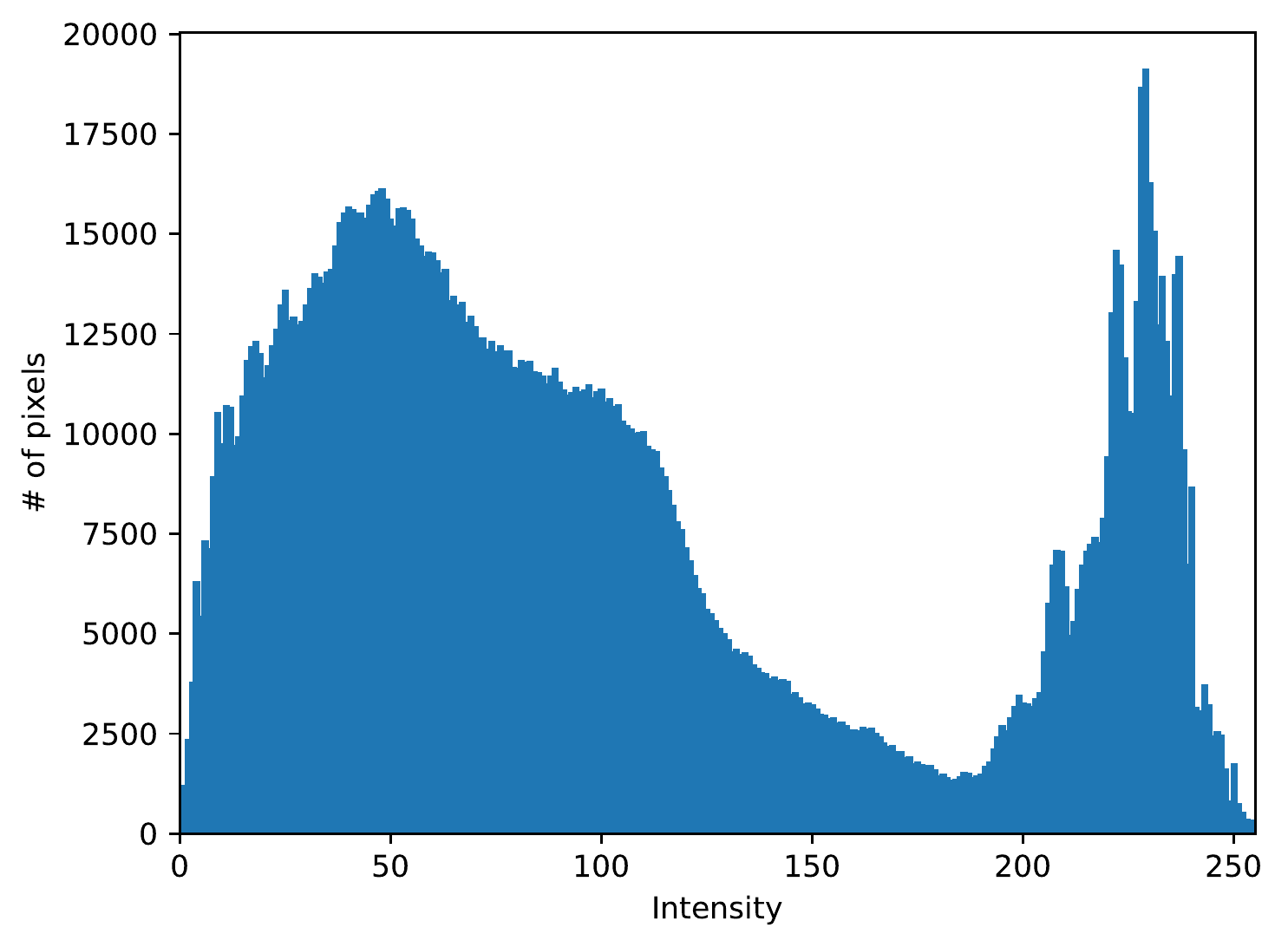}
    \end{center}
    \caption{Without color \\correction\\ $VMAF = 51.005$, \\$SSIM = 0.715$}
    \label{fig:crowd_run}
    \end{subfigure}
    \hspace*{\fill} 
    \begin{subfigure}{0.23\textwidth}
    \begin{center}
    \includegraphics[width=\linewidth]{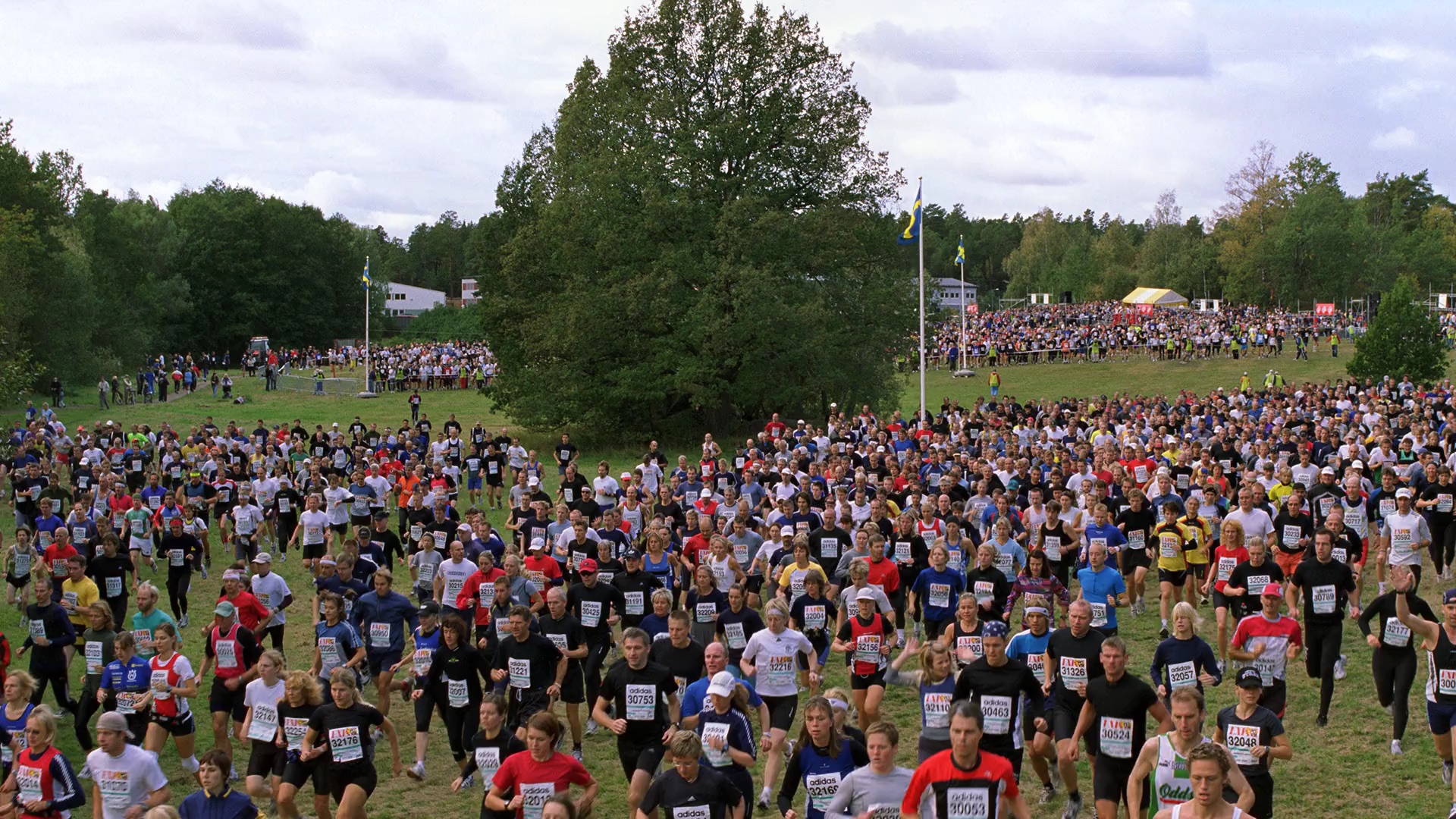}
    \\
    \includegraphics[width=\linewidth]{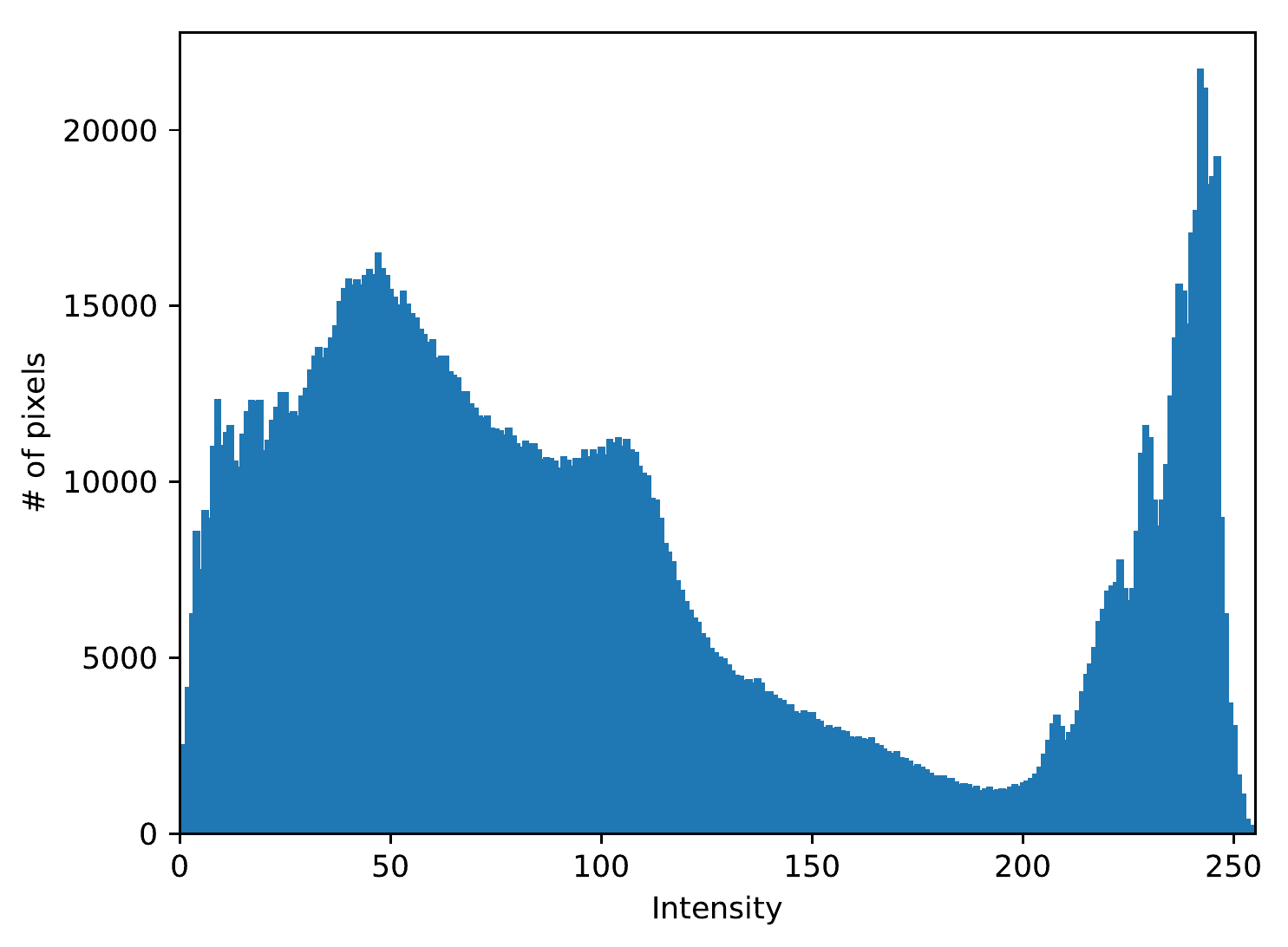}
    \end{center}
    \caption{After histogram equalization\\ $VMAF = 53.083$, \\$SSIM = 0.712$}
    \label{fig:crowd_run_adj}
    \end{subfigure}
    \caption{Frame 1 from \textit{Crowd run} video sequence and its histogram with and without color correction. Two images and their histograms look almost similar.}
\end{figure}

\begin{figure}[H]
    \includegraphics[width=\linewidth]{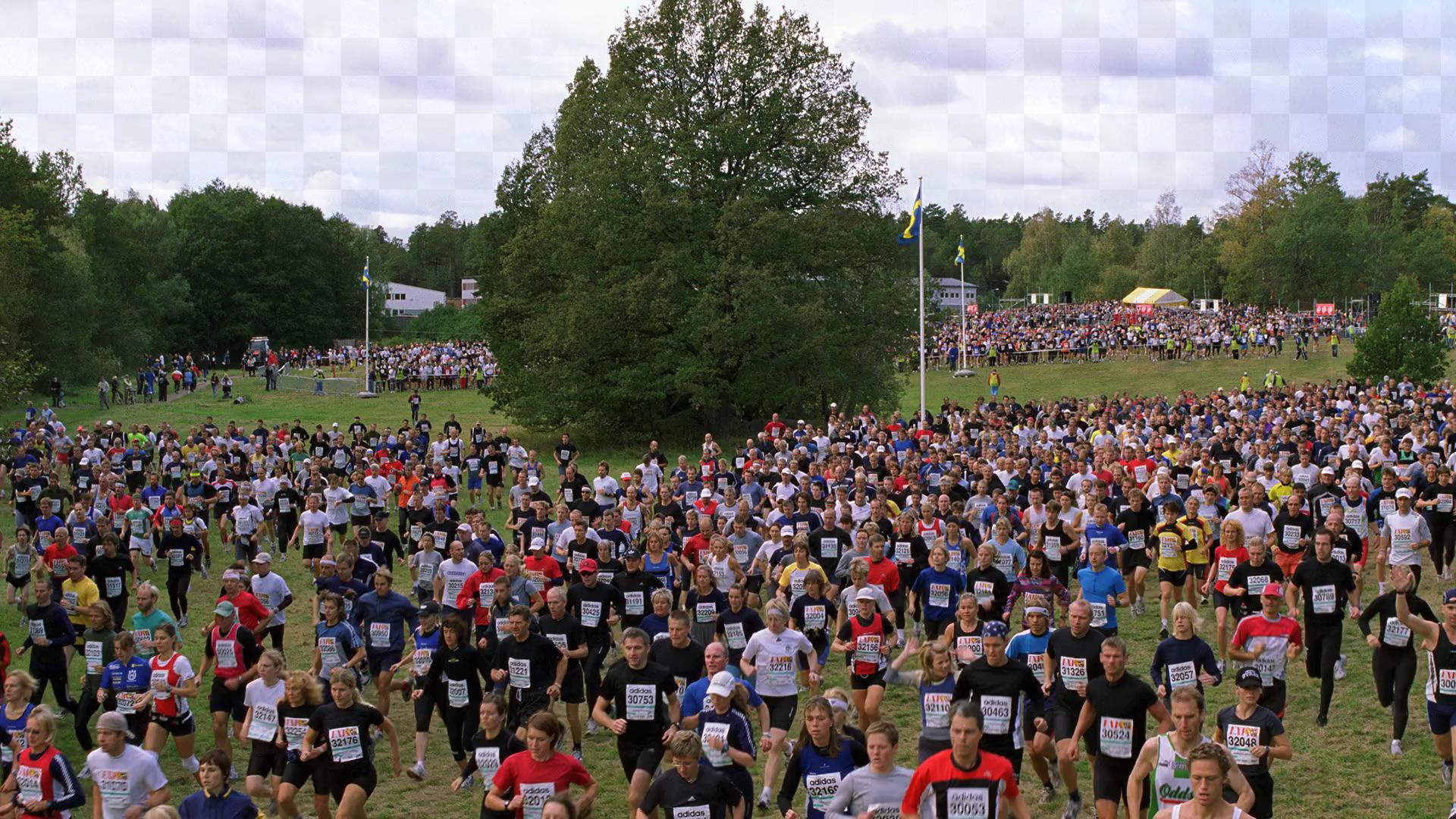}
    \caption{Checkerboard comparison of frame 1 from \textit{Crowd run} video sequence before and after distortions.}
    \label{fig:crowd_run_comparison}
\end{figure}

For \textit{Crowd run} sequence, histogram equalization with $kernel size = 8$ and $clip limit = 0.00419$ also increased VMAF (Fig.~\ref{fig:crowd_run} and Fig.~\ref{fig:crowd_run_adj}). The video is more contrasted, so the decrease in SSIM was more significant. However, tho images also look similar (Fig.~\ref{fig:crowd_run_comparison}) and have similar SSIM score, while VMAF showed better score after contrast transformation.

\textit{Red kayak} looked better according to VMAF after unsharp mask with $radius = 9.436$, $amount = 0.045$.

For \textit{Speed bag}, the following parameters of unsharp mask allowed to increase VMAF greatly without influencing SSIM: $radius = 9.429$, $amount = 0.114$.

\section{Conclusion}

Video quality reference metrics are used to show the difference between original and distorted streams and are expected to take worse values when any transformations were applied to the original video. However, sometimes it is possible to deceive objective metrics. In our article, we described the way to increase the values of popular full-reference metric VMAF. If the video is \textit{not} contrasted, VMAF can be increased by color adjustments without influencing SSIM. In another case, contrasted video can also be tuned for VMAF but with little SSIM worsening.

Although VMAF has become popular and important, particularly for video codec developers and customers, there are still a number of issues in its application. This is why SSIM is used in many competitions, as well as in MSU Video-Codec Comparisons, as a main objective quality metric.

We wanted to pay attention to this problem and hope to see the progress in this are, which is likely to happen since the metric is being actively developed. Our further research will involve a subjective comparison of the proposed color adjustments to the original videos and the development of novel approaches for metric tuning.

\section{Acknowledgments}

This work was partially supported by the Russian Foundation for Basic Research under Grant 19-01-00785a.

\aboutAuthors

\end{multicols*}

\end{document}